\documentclass[journal=jacsat,manuscript=article]{achemso}

\usepackage[version=3]{mhchem}
\usepackage{amsmath}

\newcommand*\Vmod{V_{\mathrm{mod}}}
\newcommand*\fmod{f_{\mathrm{mod}}}

\author{Zhu Zhang}
\email{z.zhang@uu.nl}
\author{Sanli Faez}
\affiliation{Nanophotonics, Debye Institute for Nanomaterials Science, Utrecht University, The Netherlands}

\title[Opto-iontronic iSCAT mapping]
  {Wide-Field Opto-Iontronic iSCAT Mapping of Interfacial Charging and Electrical Connectivity}

\abbreviations{iSCAT,EDL,ITO,FIB,FFT,SEM,SNR}

\begin{document}

\begin{abstract}
Spatial variations in electrical connectivity and interfacial ion accumulation can strongly influence electrochemical performance, yet these properties are difficult to visualize directly with conventional ensemble measurements. Here, we introduce potential-modulated opto-iontronic microscopy, an interferometric scattering microscopy (iSCAT) approach for wide-field imaging of electric-double-layer (EDL) dynamics at nanostructured electrodes. Sinusoidal potentials were applied to focused-ion-beam-fabricated indium tin oxide (ITO) nanoholes, and the optical response was extracted at the modulation frequency by Fourier demodulation. The optical modulation amplitude increased approximately linearly with modulation voltage above a low-voltage roll-off and decreased with increasing frequency, consistent with kinetically limited interfacial charging. We then mapped the potential-synchronized optical amplitude across patterned ITO electrodes. Electrically isolated blocks exhibited strongly suppressed modulation signals, whereas electrically connected and partially milled nanogrid structures showed pronounced responses. These results demonstrate label-free optical mapping of local charging dynamics and electrical connectivity at heterogeneous electrochemical interfaces using a commercially available iSCAT platform.
\end{abstract}

\section{Introduction}

Electrochemical performance is governed not only by the average properties of an electrode but also by spatial variations in electrical connectivity, surface morphology, and interfacial ion accumulation \cite{bentley_scanning_2017,wu_understanding_2022, wei_accessing_2020}. In particular, heterogeneous electrical coupling between nanoscale objects and their supporting electrode can produce substantial variations in their apparent electrochemical activity \cite{wei_accessing_2020}. Local disconnection, structural defects, and nanoscale confinement can therefore lead to heterogeneous charging and electrochemical behavior that is difficult to resolve using conventional ensemble measurements \cite{chen_heterogeneous_2014}. Conventional voltammetry and electrochemical impedance measurements generally report responses averaged over the electrode surface and can consequently obscure such local heterogeneity \cite{huang_local_2011,bentley_scanning_2017}. Spatially resolved optical approaches can complement these measurements by providing non-invasive visualization of local electrochemical processes \cite{wang_optical_2019}.

Interferometric scattering microscopy (iSCAT) is a label--free optical technique that detects the interference between light scattered by an object and a reference field reflected from an interface. Its high sensitivity has enabled detection and tracking of nanoscale objects and biomolecular assemblies \cite{lindfors_detection_2004,kukura_high-speed_2009,piliarik_direct_2014,ginsberg_interferometric_2025,hsieh_tracking_2014,tala_pseudomonas_2019}. Scattering-based microscopies have subsequently been extended to materials and electrochemical systems\cite{grobmeyer_visualizing_2026}, including the visualization of charge and energy transport\cite{delor_imaging_2020,sung_long-range_2020}, lithium-ion dynamics\cite{merryweather_operando_2021},  particle degradation\cite{merryweather_operando_2022}, ion channel activity\cite{li_electrochemically_2025}, nanoparticle nucleation\cite{guo_real_time_2023}, andmaterial formation \cite{kowal_electrophoretic_2024,gruber_early_2024,afsahi_seeing_2025}.

Potential--modulated scattering measurements provide a direct route to isolating weak electrochemical responses from static optical backgrounds. Potentiodynamic or iontronic microscopy has been used to probe EDL charging and discharging, ion accumulation near sharp electrodes, particle trapping, and electrochemical reactions within nanostructures \cite{namink_electric-double-layer-modulation_2020,zhang_iontronic_2023,dogru_yuksel_laser-patterned_2025,zhang_optical_2026}. Related wide-field and plasmonic approaches have visualized ion redistribution and investigated the optical contributions of cations and anions during interfacial charging and chemical reactions\cite{utterback_operando_2023,regules-medel_untangling_2025,liu_electrochemical_2018,niu_determining_2022}. These studies establish that potential-driven ion redistribution can generate measurable refractive-index and scattering changes near an electrode.

However, three questions remain central for applying this concept to heterogeneous nanoscale electrodes. First, can potential--modulated iSCAT detect EDL charging in confined nanostructures using a broadly accessible commercial microscope? Second, what voltage and frequency ranges define the measurable optical response? Third, can spatial amplitude maps distinguish electrically connected regions from electrically isolated structures, even when both are visible in a conventional static image?

Here, we address these questions using FIB--fabricated ITO nanohole arrays and two-dimensional patterned electrodes. We first establish potential-synchronized optical contrast from individual nanoholes and isolate the response in the Fourier domain. We then determine how the optical amplitude varies with modulation voltage and frequency. Finally, we map the potential-synchronized optical amplitude across isolated and connected ITO patterns. The combined measurements show that dynamic optical contrast reports both local interfacial charging and electrical connectivity, providing a label-free route to identifying heterogeneous electrochemical behavior over a wide field of view.

\section{Results and Discussion}

\subsection{Potential-Modulated iSCAT Imaging of Nanohole Electrodes}

\begin{figure}[htbp]
\centering
\includegraphics[width=\columnwidth]{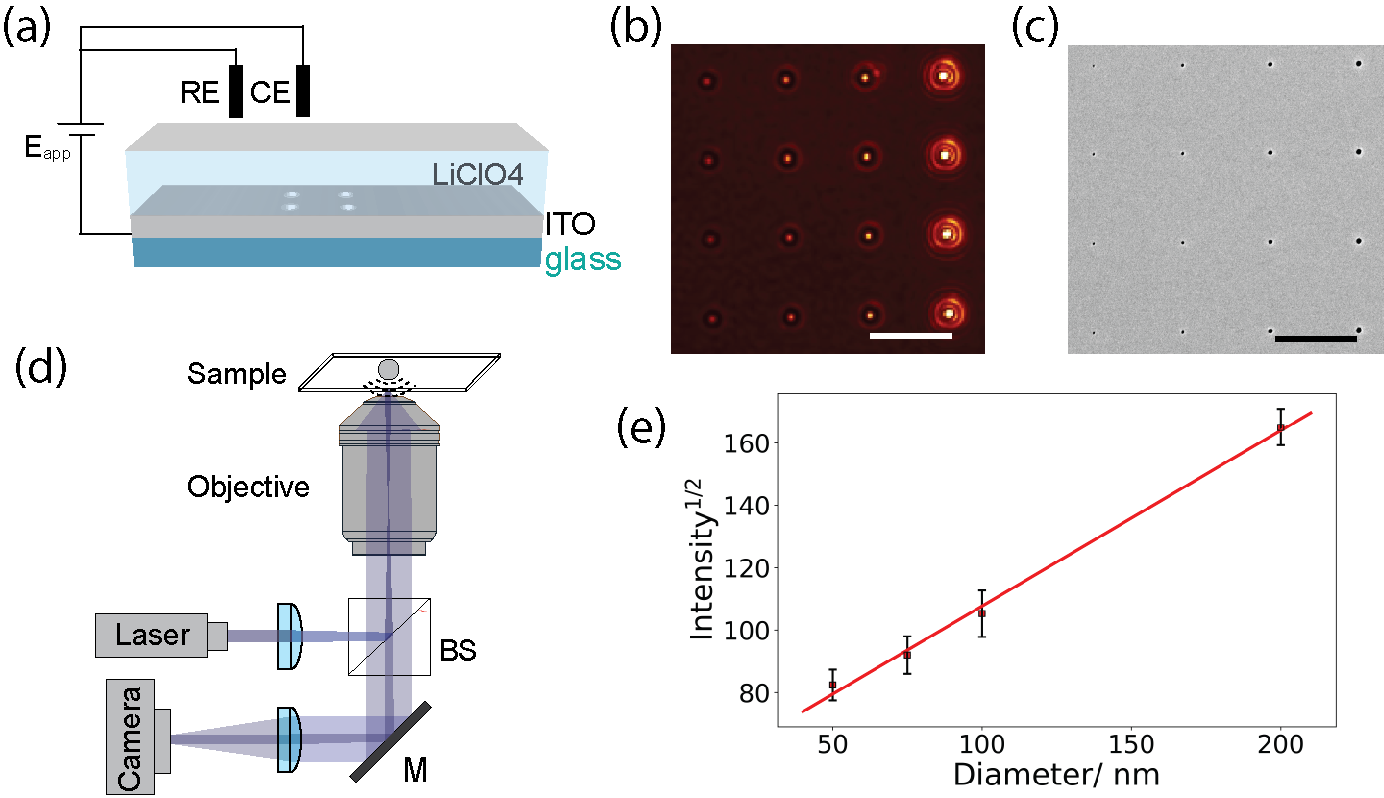}
\caption{Experimental configuration and nanohole characterization. (a) Schematic of the three-electrode electrochemical cell containing the ITO nanohole working electrode, Ag/AgCl reference electrode (RE), Pt counter electrode (CE), and 100~mM \ce{LiClO4} electrolyte. (b) iSCAT image of the nanohole array. (c) Corresponding SEM image of the FIB-fabricated array. (d) Schematic of the iSCAT optical configuration. BS, beam splitter; M, mirror. (e) Square root of the measured nanohole intensity as a function of nominal nanohole diameter. The line is a linear fit.  Error bars represent the standard deviation of the 4 nanoholes at each diameter. Scale bars: $6$~$\mu m$.}
\label{fig_setup}
\end{figure}

We implemented potential-modulated opto-iontronic imaging on a commercial OneMP interferometric scattering microscope (Refeyn, UK). The electrochemical and optical configuration is summarized in Figure~\ref{fig_setup}. A wide-field light illuminates the ITO--electrolyte interface, and the objective collects both the reference field reflected from the interface and the field scattered by the nanostructures. Their interference is recorded by a camera, allowing small changes in the scattering response to be detected as time--dependent intensity variations. The illumination wavelength is $525$~nm. Nanohole arrays with nominal diameters from $50$ to $200$~nm were fabricated in ITO by focused ion beam (FIB) milling. The nanoholes provide spatially localized scattering features and confined electrode--electrolyte interfaces with a high surface-to-volume ratio. The patterned ITO served as the working electrode in a three--electrode electrochemical cell containing an Ag/AgCl reference electrode, a Pt counter electrode, and 100~mM 
\ce{LiClO4} electrolyte. The cell dimensions are $5$~mm in diameter and $2$~mm in height.

The iSCAT image in Figure~\ref{fig_setup}b shows the nanohole array, and the corresponding SEM image in Figure~\ref{fig_setup}c confirms the fabricated structures. Each array contains nanoholes with diameters of $50$~nm, $75$~nm, $100$~nm, and $200$~nm, and there are $4$ nanoholes for each size. Figure~\ref{fig_setup}(e) shows that the square root of the measured nanohole intensity increases approximately linearly with nanohole diameter.Because all nanoholes were milled to approximately the same depth \(H\), is constant for all nanoholes. Under the simplifying assumption that the measured optical intensity scales approximately with nanohole volume over the investigated size range, \(I \propto V_{\mathrm{hole}} = \frac{\pi}{4}D^{2}H\), where \(D\) is the nanohole diameter and \(H\) is its depth. Because \(H\) is constant, \(I \propto D^{2}\), and giving \(\sqrt{I} \propto D\), consistent with the observed trend. We note, however, that the absolute iSCAT response also depends on the complex scattered field, reference field, focus, and interferometric phase, so this scaling should be regarded as an empirical geometric interpretation rather than a complete scattering model.

\begin{figure}[htbp]
\centering
\includegraphics[width=\columnwidth]{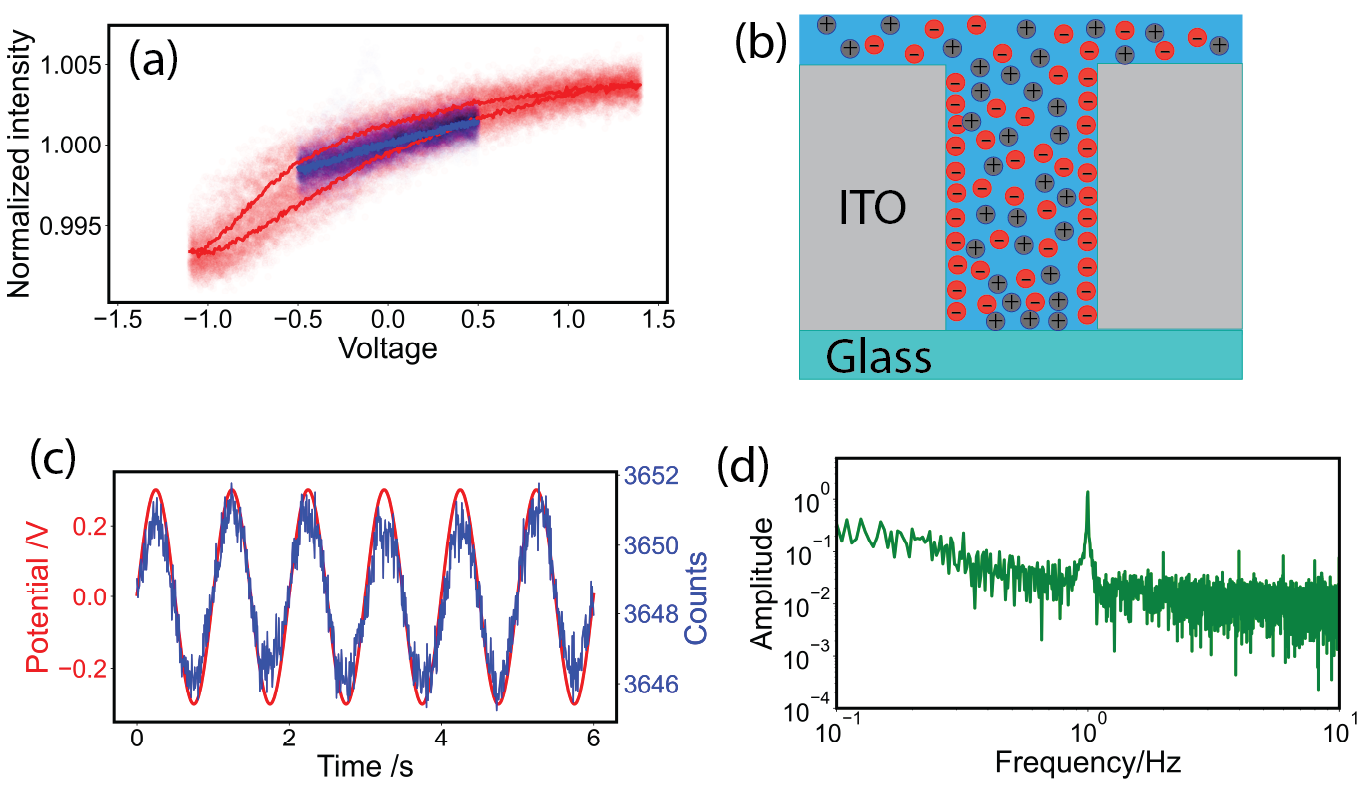}
\caption{Potential-synchronized optical response of an ITO nanohole. (a) Normalized interferometric intensity of a single nanohole during a cyclic potential scan. The highlighted interval indicates the potential window used for modulation experiments. (b) Schematic illustration of ion redistribution within an ITO nanohole under positive applied potential. (c) Applied sinusoidal potential (red) and corresponding nanohole intensity trace (blue) for $\Vmod=0.2$~V and $\fmod=1$~Hz. (d) Fourier amplitude of the intensity trace, showing a peak at the applied modulation frequency. }
\label{fig_fft_results}
\end{figure}

To identify a potential range dominated by interfacial charging rather than pronounced Faradaic reactions, we first measured the optical response of a nanohole during a cyclic potential scan (Figure~\ref{fig_fft_results}a). The normalized nanohole intensity increased as the potential was scanned from approximately $-1.1$ to $1.4$~V and returned toward its initial level on the reverse scan with a scan rate $1250$~mV/s. To minimize contributions from oxygen reduction, ITO oxidation, and other Faradaic processes, subsequent modulation measurements were restricted to the highlighted potential window from $-500$~mV to $500$~mV, in which no pronounced Faradaic features were observed in the corresponding cyclic voltammogram (Figure S1).

We therefore applied a sinusoidal potential with amplitude $V_{mod} = 0.2$ V and frequency $f_{mod}=1$~Hz, such that the potential excursion remained within the predominantly non-Faradaic window identified above. The optical intensity of an individual nanohole followed the applied waveform (Figure~\ref{fig_fft_results}c), indicating a potential-synchronized change in the local optical response. During negative polarization, cations such as \ce{Li+} preferentially screen the charged ITO surface, whereas anions such as \ce{ClO4-} contribute more strongly during positive polarization. The resulting redistribution of ions and solvent near the interface changes the local refractive index and therefore the interferometric scattering signal.

For each nanohole, we extracted the intensity as a function of time and calculated its Fourier transform. Figure~\ref{fig_fft_results}d shows a distinct spectral peak at the applied modulation frequency. Fourier-domain analysis suppresses static structure and separates the potential-synchronized response from broadband fluctuations. The amplitude of the Fourier coefficient at the fundamental modulation frequency was used as the optical modulation amplitude.

The relationship between ionic composition and refractive index can be described using the Lorentz--Lorenz relation. For a single component,

\begin{equation}
R = \frac{n^{2}-1}{n^{2}+2}\frac{1}{c}
  = \frac{4\pi}{3}N_{\mathrm{A}}\alpha,
\label{eq:lorentz-lorenz}
\end{equation}

where $n$ is the refractive index, $c$ is the molar concentration, $N_{\mathrm{A}}$ is Avogadro's constant, $\alpha$ is the molecular polarizability, and $R$ is the molar refractivity \cite{p_molar_1989}. For a mixture of solvent and ionic species, an additive approximation gives

\begin{equation}
\frac{n^{2}-1}{n^{2}+2}=\sum_i R_i c_i,
\label{eq:mixing}
\end{equation}

where $R_i$ and $c_i$ denote the molar refractivity and concentration of component $i$ \cite{pauling_theoretical_1927,bouteloup_improved_2018,namink_electric-double-layer-modulation_2020, regules-medel_untangling_2025}. Equations~\ref{eq:lorentz-lorenz} and \ref{eq:mixing} motivate the sensitivity of the optical signal to ion redistribution. However, the measured iSCAT amplitude also depends on nanohole geometry, focus, and interferometric phase; therefore, the present data are interpreted as a relative potential-driven optical response rather than an absolute ion-concentration measurement.

\subsection{Voltage Dependence and Low-Voltage Roll-Off}

\begin{figure}[htbp]
\centering
\includegraphics[width=\columnwidth]{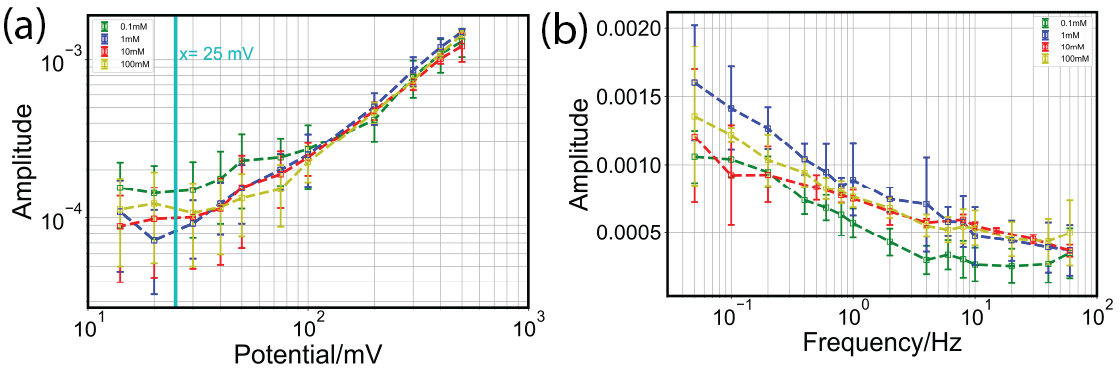}
\caption{Voltage and frequency dependence of the nanohole response. (a) Optical modulation amplitude as a function of potential-modulation amplitude for the indicated \ce{LiClO4} concentrations. The vertical line marks the room-temperature thermal-voltage scale ($k_{\mathrm{B}}T/e\approx25$~mV) as a reference guide. Dashed curves are guides to the eye. (b) Optical modulation amplitude as a function of modulation frequency from 0.05--80~Hz, with the modulation potential at $V_{mod} = $ 0.2~V, at the nanohole with a diameter of 100~nm.}
\label{fig_scan_results}
\end{figure}

Having established a potential-synchronized optical signal, we next quantified the dependence of the modulation amplitude on the applied voltage. The modulation amplitude was varied from $10$ to $500$~mV for several \ce{LiClO4} concentrations (Figure~\ref{fig_scan_results}a). At modulation amplitudes above approximately $100$~mV, the optical response increased approximately linearly with $V_{mod}$, indicating that the potential--driven change in interfacial optical properties increased with the electrical excitation. At lower voltages, the response deviated from the high--voltage trend and approached a plateau near the smallest applied amplitudes.

The thermal voltage at room temperature is $k_{\mathrm{B}}T/e\approx24.8$~mV. The observed roll-off occurs as the applied modulation approaches this voltage scale. This comparison is physically suggestive because the applied electrostatic energy approaches the thermal--energy scale. Nevertheless, the present data do not by themselves demonstrate that the instrument is thermal--noise limited. Instrumental noise, mechanical drift, camera noise, finite Fourier resolution, and nonlinear electrochemical charging may also contribute. We therefore describe the result as a low--voltage roll--off near the thermal--voltage scale rather than as proof of a thermal--noise limit.

\subsection{Frequency Response of Interfacial Charging}

The modulation frequency was varied while the voltage amplitude was held at $V_{mod}= 200$~mV. As shown in Figure~\ref{fig_scan_results}b, the optical amplitude decreased with increasing frequency. At low frequency, the interfacial ion distribution has sufficient time to approach its quasi-equilibrium state during each half--cycle. At higher frequency, the EDL cannot fully follow the applied potential, resulting in an attenuated optical response.

An approximate EDL charging time can be estimated as $\tau_{\mathrm{RC}}=\lambda_{\mathrm{D}}L/D\approx5$~ms, with Debye length $\lambda_{\mathrm{D}}\simeq1$~nm, system length $L\simeq10$~mm, and diffusion coefficient $D\simeq2\times10^{-9}$~m$^2$~s$^{-1}$\cite{zhang_optical_2026}. 

\subsection{Wide-Field Mapping of Local Electrical Connectivity}

\begin{figure}[htbp]
\centering
\includegraphics[width=\columnwidth]{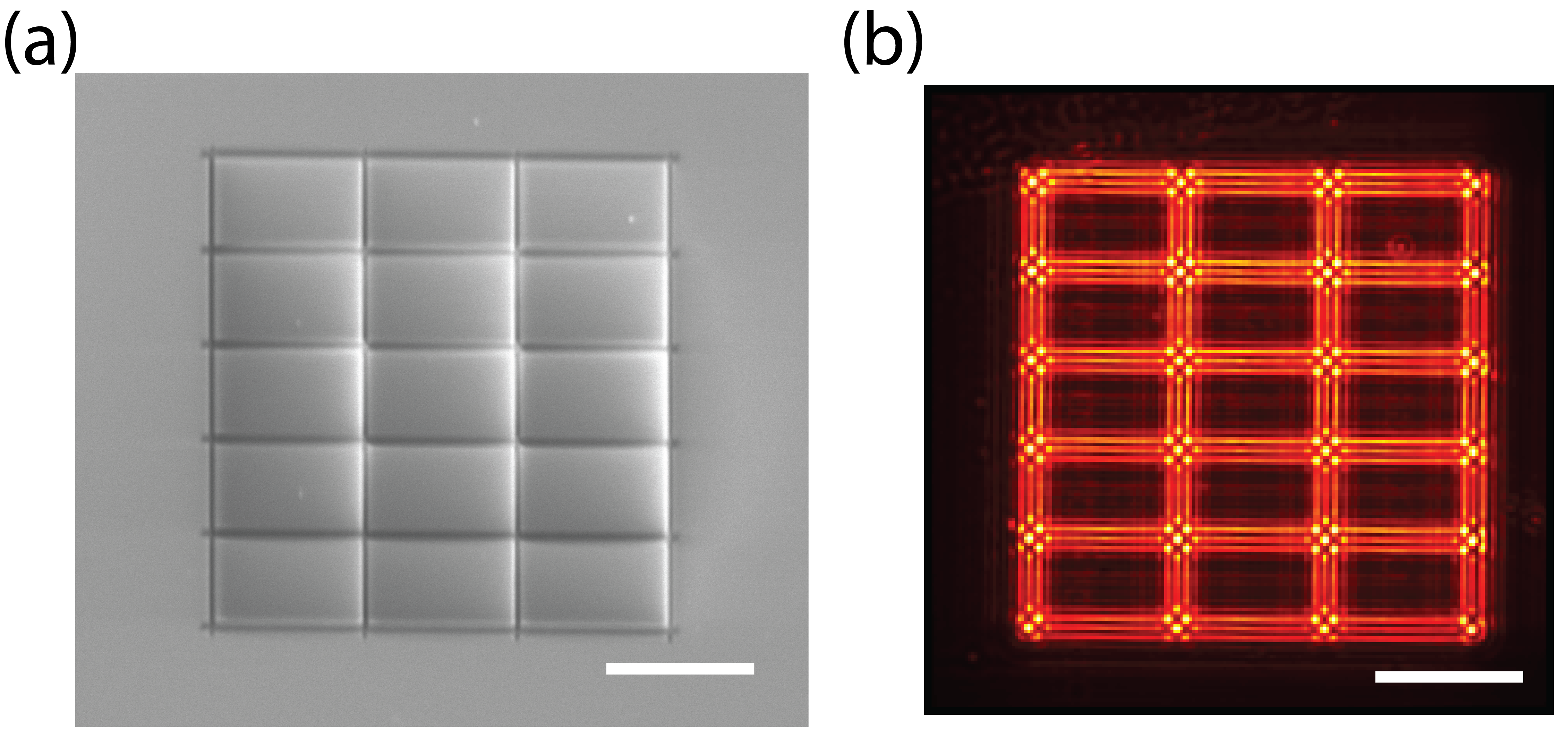}
\caption{Structural characterization of the isolated ITO block array. (a) SEM image of the FIB-patterned $3\times4$ array. (b) Corresponding static iSCAT image. The inner blocks were designed to be electrically isolated from the surrounding continuous ITO. Scale bars: 5 $\mu $m.}
\label{fig_2D_electrode}
\end{figure}

We next extended the measurement from individual nanoholes to two--dimensional patterned ITO electrodes. The central hypothesis is that if the modulation signal reports local interfacial charging, electrically connected regions should follow the applied potential, whereas electrically isolated regions should exhibit strongly suppressed responses. To test this hypothesis, we fabricated a $3\times4$ array of rectangular ITO blocks by milling through the conductive ITO layer around each block. The SEM and static iSCAT images are shown in Figure~\ref{fig_2D_electrode}. The surrounding continuous ITO remained connected to the electrical contact, whereas the inner blocks were designed to be electrically isolated.

The static iSCAT image clearly resolves the patterned structure, but static contrast alone does not identify whether a region is electrically connected. Dynamic amplitude maps provide this additional information. For each camera pixel, the temporal signal was Fourier transformed and the complex coefficient at $f_{mod}$ was extracted. The resulting amplitude map reports the strength of the potential--synchronized response.

\begin{figure}[htbp]
\centering
\includegraphics[width=\columnwidth]{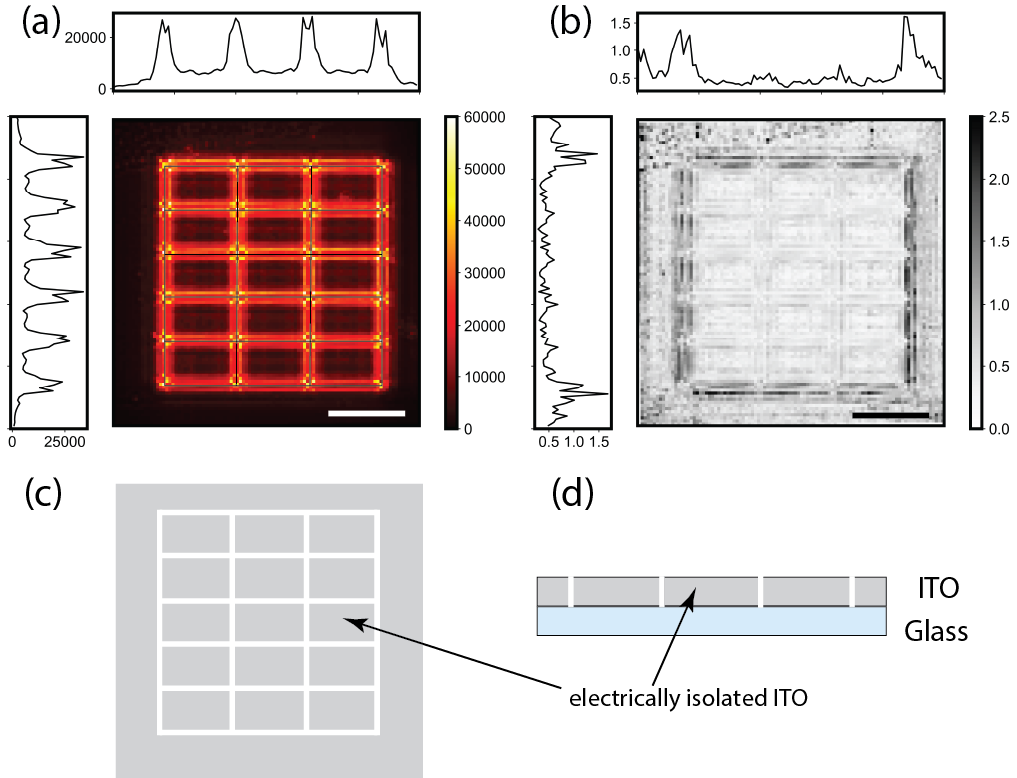}
\caption{Dynamic optical response of electrically isolated ITO blocks. (a) Static iSCAT image of the FIB-patterned ITO array with corresponding horizontal and vertical intensity profiles. (b) Fourier-demodulated optical modulation-amplitude map acquired at $V_{\mathrm{mod}}=0.2$ V and $f_{\mathrm{mod}}=4$ Hz. The interiors of the electrically isolated ITO blocks exhibit strongly suppressed modulation amplitudes compared with the electrically connected surrounding ITO. The modulation amplitude is reported in arbitrary units corresponding to the magnitude of the Fourier coefficient at $f_{\mathrm{mod}}$. (c) Top-view schematic of the patterned ITO array, showing the electrically isolated blocks separated from the surrounding ITO by FIB--milled trenches. (d) Cross-sectional schematic illustrating that the trenches extend through the conductive ITO layer to the glass substrate, thereby electrically isolating the blocks. Scale bars: $5~\mu\mathrm{m}$. Acquisition time for the amplitude map: $5$~s.}
\label{fig_2D_block_results}
\end{figure}

Figure~\ref{fig_2D_block_results} compares the static intensity with the potential--synchronized amplitude. The interiors of the isolated blocks exhibit strongly suppressed modulation amplitudes relative to the surrounding connected ITO. Enhanced responses are observed at portions of the pattern edges. This edge contrast may arise from residual electrical connection, local field concentration, additional interfacial area, optical edge scattering, or a combination of these effects. The key observation is that regions with similar static morphology display different dynamic responses according to their electrical state.

\begin{figure}[htbp]
\centering
\includegraphics[width=\columnwidth]{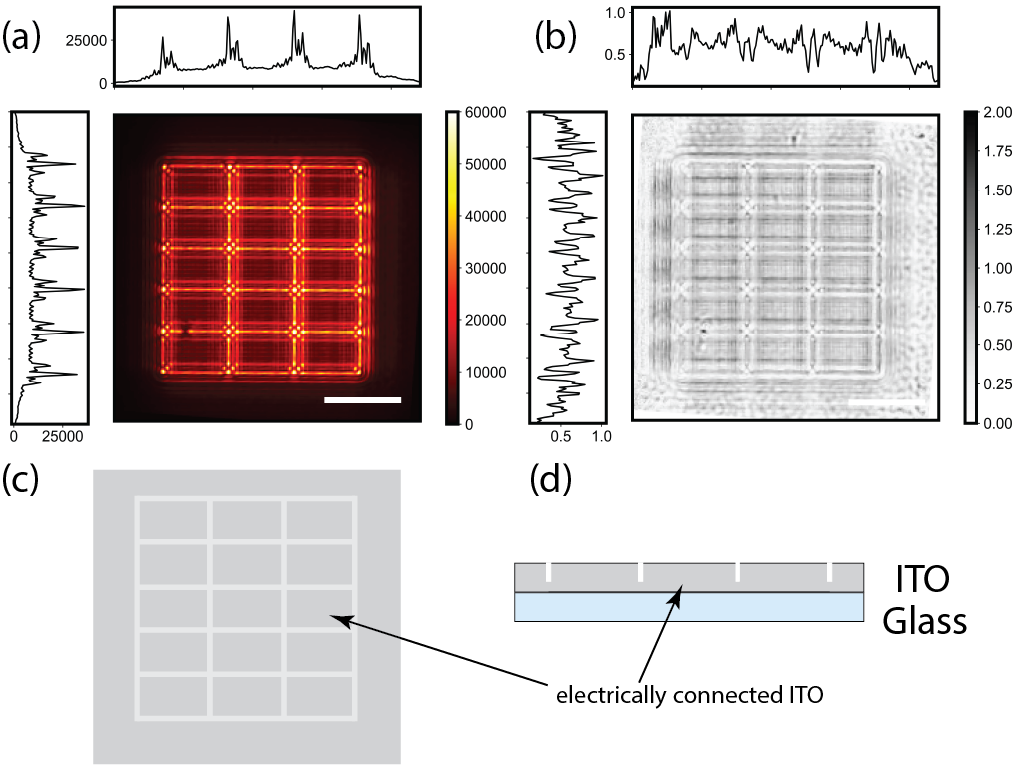}
\caption{Dynamic optical response of an electrically connected ITO grid. (a) Static iSCAT image of the partially milled ITO grid with corresponding horizontal and vertical intensity profiles. (b) Fourier-demodulated optical modulation-amplitude map acquired at $V_{\mathrm{mod}}=0.2$ V and $f_{\mathrm{mod}}=4$ Hz. In contrast to the electrically isolated structures in Figure 5, the connected grid lines exhibit a pronounced potential-synchronized optical response. The modulation amplitude is reported in arbitrary units corresponding to the magnitude of the Fourier coefficient at $f_{\mathrm{mod}}$. (c) Top-view schematic of the patterned ITO grid, which remains electrically connected to the surrounding electrode. (d) Cross-sectional schematic illustrating partial FIB milling, with a continuous ITO layer remaining beneath the trenches and preserving electrical connectivity. Scale bars: $5~\mu\mathrm{m}$. Acquisition time for the amplitude map: 5 s.}
\label{fig_2D_connected_blocks}
\end{figure}

To provide a structural control, we fabricated a similar grid pattern without milling fully through the ITO layer. These partially milled lines remained electrically connected to the surrounding electrode. The static iSCAT image in Figure~\ref{fig_2D_connected_blocks}a shows strong scattering from the grid lines, which had a nominal width of approximately $100$~nm and depth of approximately $50$~nm. Under potential modulation, the grid pattern produced a clear dynamic response (Figure~\ref{fig_2D_connected_blocks}b), in contrast to the suppressed interiors of the fully isolated blocks.
The electrical state of the patterned ITO structures was additionally assessed using ion-beam-induced charging contrast during FIB imaging (Figure S2). Under the imaging conditions used here, electrically isolated regions exhibited pronounced dark charging contrast, whereas structures connected to the surrounding conductive ITO maintained a substantially more stable secondary-electron signal. This contrast arises from differences in beam-induced charging and the resulting modification of ion-induced secondary-electron emission and collection. The FIB images therefore provide qualitative supporting evidence for the electrical isolation or continuity of the patterned regions.

We also examined an ITO region containing irregular surface structures visible in the static iSCAT image (Figure S3 in SI). The demodulated amplitude map revealed enhanced and spatially structured responses near these features, demonstrating that the method is sensitive to electrochemical heterogeneity outside deliberately patterned arrays. Together, the isolated and connected patterns demonstrate that potential-modulated optical contrast distinguishes local electrical continuity from static morphology.

\section{Conclusion}

In this work, we implemented potential-modulated opto-iontronic microscopy on a commercial iSCAT platform and used it to visualize EDL charging at nanostructured ITO electrodes. FIB--fabricated nanoholes produced potential--synchronized optical signals that were isolated at the applied frequency by Fourier demodulation. The optical amplitude increased approximately linearly with modulation voltage above a low-voltage roll-off and decreased with increasing frequency, consistent with finite--rate interfacial charging.

Extending the method to two-dimensional patterned electrodes revealed its central spatial capability. Electrically isolated ITO blocks exhibited strongly suppressed modulation signals, whereas connected edges and partially milled grid structures showed enhanced responses. The comparison demonstrates that dynamic amplitude maps contain information about local electrical continuity that is not available from static iSCAT images alone.

Potential--modulated iSCAT therefore provides a label--free optical readout of local interfacial charging and electrical connectivity over a wide field of view, offering a route toward identifying inactive regions, weak contacts, and heterogeneous charging in nanoscale electrode architectures.

\section{Methods}

\subsection{ITO Substrate Preparation and FIB Fabrication}

ITO-coated glass substrates with dimensions of $22\times40$~mm, nominal substrate thickness \#1, and sheet resistance of 70--100~$\Omega\,\mathrm{sq}^{-1}$ were purchased from SPI Supplies. The substrates were rinsed sequentially with deionized water, ethanol, and isopropanol, and then dried with \ce{N2}. Nanoholes and grid structures were fabricated using an FEI dual-beam FIB/SEM system. The surface was first focused in SEM mode to minimize ion-beam exposure. The stage was then tilted by $52^{\circ}$ to orient the sample normal to the ion beam. After low--dose focusing in FIB mode, nanoholes and block patterns were milled using a beam current of 27~pA at 30~kV acceleration voltage, with a dwell time of 50~$\mu$s and a $75$~nm Z--depth.

Nanoholes had nominal diameters between 50 and 200~nm. For the isolated-block samples, trenches were milled through the conductive ITO layer around a $3\times4$ block array. For the connected-grid samples, the milling depth was limited so that the patterned lines remained electrically connected to the surrounding ITO. Ion-beam image contrast was used during fabrication as an initial indicator of isolation, which is shown in Figure S2(d).

\subsection{Electrochemical Cell and Potential Modulation}

A silicone gasket was attached to the patterned ITO substrate to define the electrochemical cell with a diameter of $5$~mm and thickness of $2$~mm. Copper tape contacted the ITO outside the liquid well. Approximately 20~$\mu$L of $100$~mM \ce{LiClO4} solution was added to the cell. A Pt wire with a diameter of $0.25$~mm served as the counter electrode, and an Ag/AgCl wire with a diameter of $0.2$~mm served as the reference electrode. The wires were positioned approximately 10~mm above the ITO surface. The three electrodes were connected to a BioLogic SP-200 potentiostat. For modulation measurements, a sinusoidal potential
$V(t) = V_{\mathrm{offset}} + V_{\mathrm{mod}} \sin(2\pi f_{\mathrm{mod}} t)$
was applied to the ITO working electrode.  The acquisition duration was set to $120$~s for all the frequencies above $1$~Hz, and set to $240$~s for frequencies below $1$~Hz.

\subsection{iSCAT Optical Measurements}

Data were acquired on a OneMP interferometric scattering microscope (Refeyn, UK) using circularly polarized illumination at wavelength 525~nm with an objective NA of $1.49$ and a magnification of $100$$\times$. The instrument recorded wide-field movies of the ITO--electrolyte interface at a frame rate of $160$~$fps$.

\subsection{Nanohole Time-Trace Analysis}

Raw movies were processed using custom Python routines. Nanoholes were identified in a reference frame either manually or using the \textit{trackpy} package. For each nanohole, the intensity was integrated over a fixed region of interest in every frame to generate a time trace $I(t)$. 
The discrete Fourier transform of each trace was calculated to determine the optical modulation amplitude. The applied modulation frequency is therefore referred to as the fundamental frequency.

\subsection{Pixelwise Amplitude Mapping}

For two-dimensional maps, the Fourier transform was calculated along the time axis independently for every image pixel. The amplitude of the complex Fourier coefficient at $f_{mod}$ was used to construct the modulation-amplitude maps shown here. Because the Fourier coefficient directly contains the response at the selected frequency, an inverse Fourier transform was not required for constructing these amplitude maps.

\subsection{SEM and FIB Characterization}

Nanohole arrays and patterned ITO structures were characterized using the SEM column of the FEI dual-beam system. After FIB fabrication, the stage was returned to $0^{\circ}$ for imaging. Micrographs were acquired at an accelerating voltage of $5$~kV and nominal magnification of $10{,}000\times$, with a working distance of $4.0$~mm. Ion-beam-induced charging contrast was used as an additional qualitative indicator of electrical isolation. Under the FIB imaging conditions used here, isolated ITO regions exhibited pronounced dark charging contrast, whereas electrically connected structures exhibited substantially weaker charging contrast.

\begin{acknowledgement}
The authors thank Arjan Driessen, Jan Bonne Aans, and Peter van den Beld for technical support. Z.~Z. acknowledges funding from the Dutch Research Council (NWO-XS, OCENW.XS24.3.314) and the Utrecht University Pathways to Sustainability programme. This research was supported by Refeyn Ltd.
\end{acknowledgement}


\bibliography{achemso-demo}

\end{document}


\pagestyle{fancy}
\thispagestyle{plain}
\fancypagestyle{plain}{
\renewcommand{\headrulewidth}{0pt}
}

\makeFNbottom
\makeatletter
\renewcommand\LARGE{\@setfontsize\LARGE{15pt}{17}}
\renewcommand\Large{\@setfontsize\Large{12pt}{14}}
\renewcommand\large{\@setfontsize\large{10pt}{12}}
\renewcommand\footnotesize{\@setfontsize\footnotesize{7pt}{10}}
\makeatother

\renewcommand{\thefootnote}{\fnsymbol{footnote}}
\renewcommand\footnoterule{\vspace*{1pt}%
\color{cream}\hrule width 3.5in height 0.4pt \color{black}\vspace*{5pt}} 
\setcounter{secnumdepth}{5}

\makeatletter 
\renewcommand\@biblabel[1]{#1}            
\renewcommand\@makefntext[1]%
{\noindent\makebox[0pt][r]{\@thefnmark\,}#1}
\makeatother 
\renewcommand{\figurename}{\small{Fig.}~}
\sectionfont{\sffamily\Large}
\subsectionfont{\normalsize}
\subsubsectionfont{\bf}
\setstretch{1.125} 
\setlength{\skip\footins}{0.8cm}
\setlength{\footnotesep}{0.25cm}
\setlength{\jot}{10pt}
\titlespacing*{\section}{0pt}{4pt}{4pt}
\titlespacing*{\subsection}{0pt}{15pt}{1pt}

\fancyfoot{}
\fancyfoot[LO,RE]{\vspace{-7.1pt}\includegraphics[height=9pt]{head_foot/LF}}
\fancyfoot[CO]{\vspace{-7.1pt}\hspace{13.2cm}\includegraphics{head_foot/RF}}
\fancyfoot[CE]{\vspace{-7.2pt}\hspace{-14.2cm}\includegraphics{head_foot/RF}}
\fancyfoot[RO]{\footnotesize{\sffamily{1--\pageref{LastPage} ~\textbar  \hspace{2pt}\thepage}}}
\fancyfoot[LE]{\footnotesize{\sffamily{\thepage~\textbar\hspace{3.45cm} 1--\pageref{LastPage}}}}
\fancyhead{}
\renewcommand{\headrulewidth}{0pt} 
\renewcommand{\footrulewidth}{0pt}
\setlength{\arrayrulewidth}{1pt}
\setlength{\columnsep}{6.5mm}
\setlength\bibsep{1pt}

\makeatletter 
\newlength{\figrulesep} 
\setlength{\figrulesep}{0.5\textfloatsep} 

\newcommand{\topfigrule}{\vspace*{-1pt}%
\noindent{\color{cream}\rule[-\figrulesep]{\columnwidth}{1.5pt}} }

\newcommand{\botfigrule}{\vspace*{-2pt}%
\noindent{\color{cream}\rule[\figrulesep]{\columnwidth}{1.5pt}} }

\newcommand{\dblfigrule}{\vspace*{-1pt}%
\noindent{\color{cream}\rule[-\figrulesep]{\textwidth}{1.5pt}} }

\makeatother


\noindent\LARGE{\textbf{Supplementary Information:}} 

\noindent\LARGE{\textbf{Opto-iontronic Microscopy at the Thermal Noise Threshold$^\dag$}}\\

\noindent\large{Zhu Zhang,$^{\ast}$\textit{$^{a}$} and Sanli Faez\textit{$^{a}$}} \\

\noindent{\textit{$^{a}$~Nanophotonics, Debye Institute for Nanomaterials Science, Utrecht University, 3584CC Utrecht, The Netherlands;}}

\noindent{\textit{$^{\ast}$ E-mail:  z.zhang@uu.nl}}\\

\setcounter{figure}{0}
\renewcommand{\thefigure}{S\arabic{figure}}
\renewcommand{\figurename}{Figure}

The cyclic voltammogram of the ITO nanohole electrode was recorded from −1.1 to 1.4 V at a scan rate of 1250 mV/s in 100 mM \ce{LiClO4}. Within the potential region from 0.5~V to 0.5~V used for subsequent modulation measurements, no pronounced Faradaic peaks are observed, supporting its use as a predominantly non-Faradaic charging window.

\begin{figure}[h!]
    \centering
    \includegraphics[width=\columnwidth]{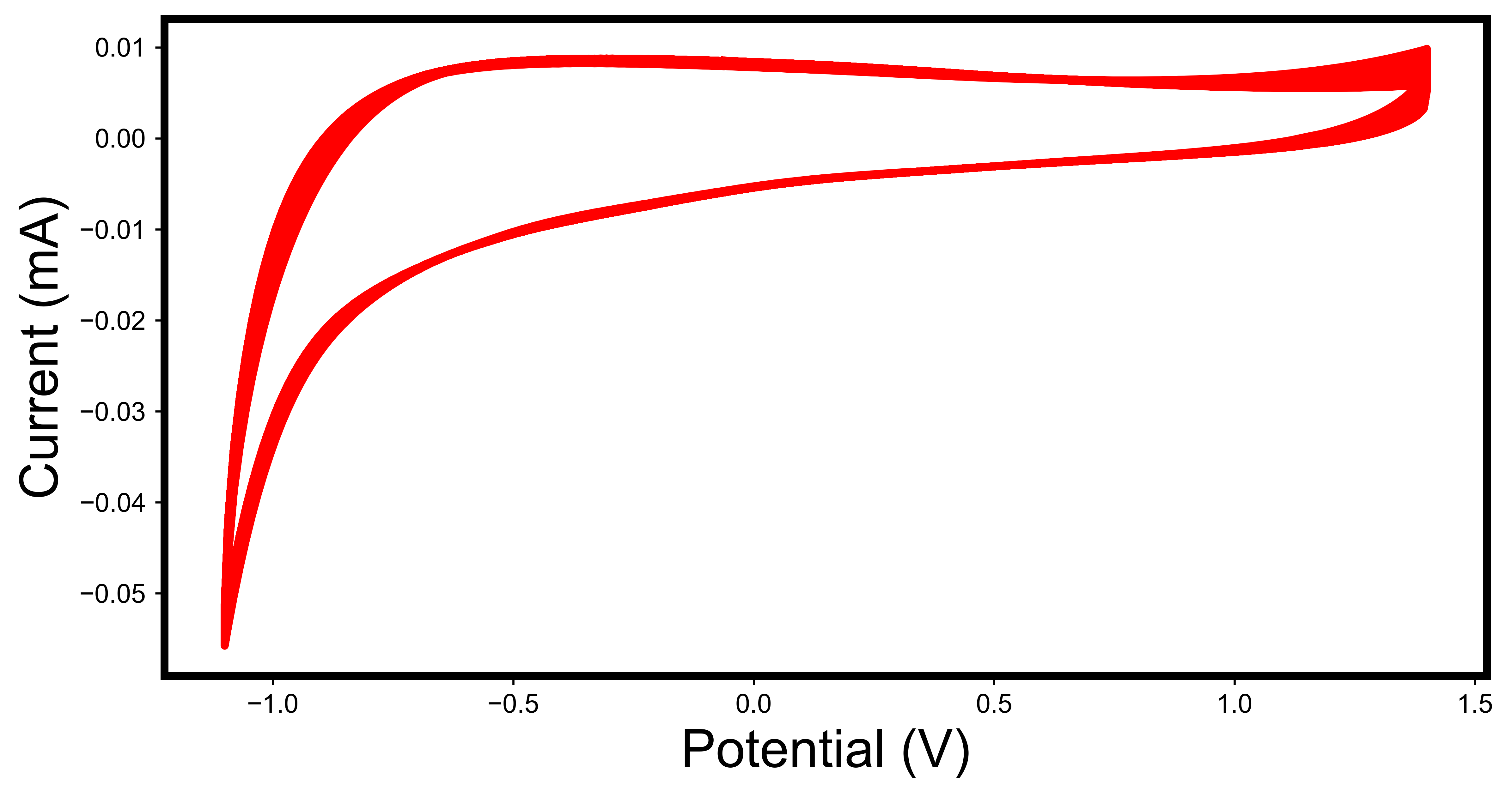}
    \caption{(a) The cyclic voltammogram of the ITO electrode with a nanoholes array. The potential is scanned from -1.1~V to 1.4~V, with scan rate of 1250~mV/s. The electrolyte is 100~mM \ce{LiClO4}. }
    \label{fig:CV}
\end{figure}

\newpage

\begin{figure}[h!]
    \centering
    \includegraphics[width=\columnwidth]{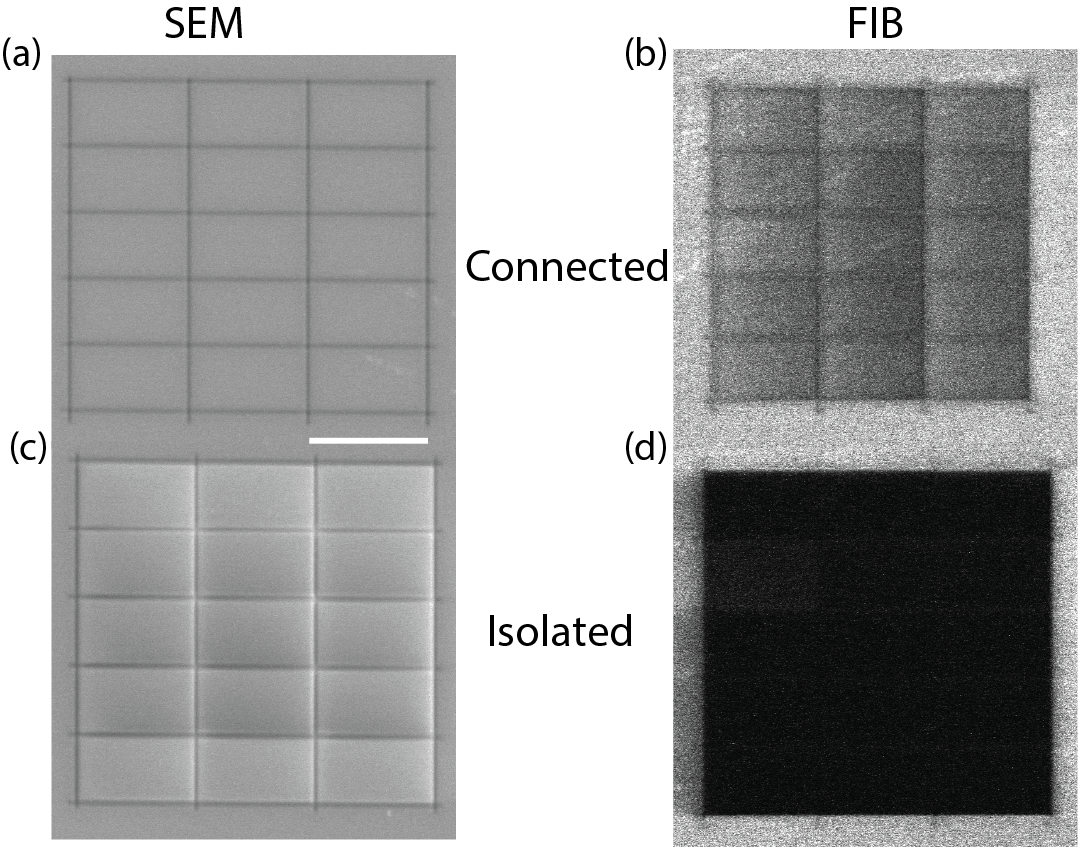}
    \caption{(a) The SEM image of the ITO blocks, which have cut the nano-grid lines, but the blocks are still connected to the ITO surface outside the ITO surface.
(b) The Focus Ion Beam (FIB) image of the same ITO blocks, which is shown in (a). (c)The SEM image of the ITO blocks, which are fully isolated from the outside the blocks. 
(c)The SEM image of the ITO blocks, which are fully isolated from the outside of the blocks. 
(d)  The corresponding FIB image of the blocks in (c).  }
    \label{fig:vsamplitude}

\end{figure}

Figure~\ref{fig:vsamplitude}(a) shows the SEM image of the partially milled ITO grid. Because the FIB-milled trenches do not extend completely through the conductive ITO layer, the patterned blocks remain electrically connected to the surrounding ITO. Consequently, no pronounced charging contrast is observed during SEM imaging, indicating that charge generated under electron-beam irradiation can be dissipated through the continuous ITO layer. Figure~\ref{fig:vsamplitude}(b) shows the corresponding FIB image of the same structure. In FIB imaging, \ce{Ga+}
 irradiation generates ion-induced secondary electrons that are collected to form the image. The connected blocks remain clearly visible and exhibit only a modest contrast relative to the surrounding ITO, consistent with the presence of an electrical pathway that limits beam-induced charging. The residual contrast may also contain contributions from local topography, FIB-induced surface modification, and variations in secondary-electron yield.

Figure~\ref{fig:vsamplitude}(c) shows the SEM image of the fully milled structure, for which the trenches extend through the conductive ITO layer and electrically isolate the inner blocks from the surrounding electrode. Pronounced charging contrast is observed during SEM imaging because the isolated regions cannot efficiently dissipate the charge generated by electron-beam irradiation. The corresponding FIB image in Figure~\ref{fig:vsamplitude}(d) shows an even stronger dark contrast from the isolated blocks. Under \ce{Ga+}
 irradiation, beam-induced charging changes the local surface potential and consequently modifies the emission and collection of ion-induced secondary electrons. Under the imaging conditions used here, this produces pronounced dark contrast for the electrically isolated blocks relative to the surrounding connected ITO. The combined SEM and FIB charging contrast therefore provides qualitative supporting evidence for the different electrical connectivity of the partially and fully milled structures.

\newpage

\begin{figure}[htbp]
\centering
\includegraphics[width=\columnwidth]{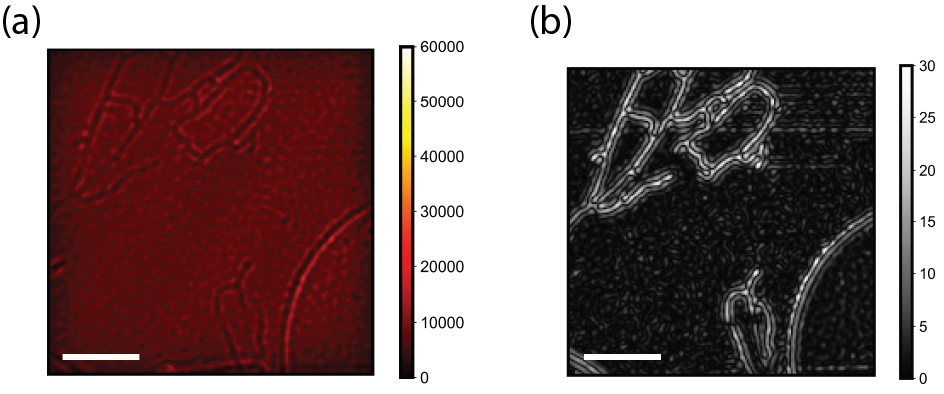}
\caption{Potential-modulated response at heterogeneous structures on an ITO surface. (a) Static iSCAT image of irregular surface features. (b) Fourier-demodulated optical amplitude at $\Vmod=0.2$~V and $\fmod= 4$~Hz. The enhanced response near the structures indicates local electrochemical and/or optical heterogeneity; their composition and conductivity have not yet been independently established. Scale bars: 5~$\mu$m.}
\label{fig_2D_dits_results}
\end{figure}

We examined an ITO region containing irregular surface structures visible in the static iSCAT image (Figure~\ref{fig_2D_dits_results}a). The demodulated amplitude map revealed enhanced and spatially structured responses near these features (Figure~\ref{fig_2D_dits_results}b), demonstrating that the method is sensitive to electrochemical heterogeneity outside deliberately patterned arrays. Their stronger modulation response may originate from local conductivity, increased surface area, altered optical scattering, topography, or modified interfacial charging. 


